\documentclass[final, 3p]{elsarticle}
\usepackage{amsmath} % assumes amsmath package installed
\usepackage{amssymb}  % assumes amsmath package installed
\usepackage{amsfonts}  % assumes amsmath package installed

\usepackage{graphics}
\usepackage[normalem]{ulem}
\usepackage{pdflscape}
\usepackage{rotating}
\usepackage{color}
\usepackage{epstopdf}

\usepackage[labelfont=bf]{caption}
\usepackage{subcaption}

\usepackage{hyperref}

\newcommand{\R}{\mathbb{R}}
\newcommand{\w}{\omega}

\newcommand{\half}{\frac{1}{2}}

\def\cale{{\cal E}}

\def\calf{{\cal F}}
\def\hal{{1 \over 2}}

\def\L2e{{\cal L}_{2e}}

\def\rea{\mathbb{R}}

\def\diag{\mbox{diag}}

\def\begequarr{\begin{eqnarray}}
\def\endequarr{\end{eqnarray}}
\def\begequarrs{\begin{eqnarray*}}
\def\endequarrs{\end{eqnarray*}}
\def\begarr{\begin{array}}
\def\endarr{\end{array}}
\def\begequ{\begin{equation}}
\def\endequ{\end{equation}}
\def\label{\label}
\def\begdes{\begin{description}}
\def\enddes{\end{description}}
\def\begenu{\begin{enumerate}}
\def\begite{\begin{itemize}}
\def\endite{\end{itemize}}
\def\endenu{\end{enumerate}}

\def\lef[{\left[\begin{array}}
\def\rig]{\end{array}\right]}

\def\begcen{\begin{center}}
\def\endcen{\end{center}}
\def\begrem{\begin{remark}\rm}
\def\endrem{\end{remark}}
\newcommand{\norm}[1]{\ensuremath{\left\| #1 \right\|}}

\def\be{\begin{equation}}
\def\ee{\end{equation}}

\newcommand{\bse}{\begin{subequations}}
\newcommand{\ese}{\end{subequations}}
\def\calh{{\cal H}}
\def\calc{{\cal C}}

\def\calm{{\cal M}}

\def\calj{{\cal J}}
\def\cals{{\cal S}}
\def\cale{{\cal E}}
\def\calf{{\cal F}}

\def\calf{{\cal F}}

\def\calj{{\cal J}}

\def\calx{{\cal X}}
\def\calu{{\cal U}}

\def\diag{\mbox{diag}}

\def\rea{\mathbb{R}}
\def\nabh{\nabla H(x)}

\def\sbar{\overline s}

\newcommand{\BP}{\noindent{\bf Proof. }}
\newcommand{\EP}{\hspace*{\fill} $\blacksquare$\smallskip\noindent}

\usepackage{amsmath,amsfonts,amssymb,amscd,latexsym,enumerate,theorem,bbm}
{\theorembodyfont{\itshape}\newtheorem{theorem}{Theorem}}
{\theorembodyfont{\itshape}\newtheorem{proposition}[theorem]{Proposition}}
{\theorembodyfont{\itshape}}
{\theorembodyfont{\itshape}\newtheorem{corollary}[theorem]{Corollary}}
{\theorembodyfont{\upshape}\newtheorem{definition}[theorem]{Definition}}
{\theorembodyfont{\itshape}}
{\theorembodyfont{\upshape}\newtheorem{remark}[theorem]{Remark}}
{\theorembodyfont{\upshape}}
{\theorembodyfont{\upshape}}
{\theorembodyfont{\upshape}\newtheorem{assumption}{Assumption}}
\usepackage[prependcaption,colorinlistoftodos]{todonotes}

\def\begali#1{\begin{align}{#1}\end{align}}
\def\begalis#1{\begin{align*}{#1}\end{align*}}

\usepackage{color}
\graphicspath{{fig/}{model/}}

\begin{document}

\begin{frontmatter}
%
%%%%%%%%%%%
\title {Conditions on Shifted Passivity of Port-Hamiltonian Systems}
%\title {Conditions for Incremental Passivity of Port-Hamiltonian Systems}
% \author{Nima Monshizadeh, Pooya Monshizadeh, Romeo Ortega, Arjan van der Schaft}
 %%%%%%%%
 %\thanks[footnoteinfo]{This paper was not presented at any IFAC 

%\thanks{Corresponding author N. Monshizadeh}
\author[CAM]{Nima Monshizadeh}\ead{n.monshizadeh@eng.cam.ac.uk}
\author[UG]{Pooya Monshizadeh}\ead{p.monshizadeh@rug.nl}
\author[SPLC]{Romeo Ortega}\ead{ortega@lss.supelec.fr}
\author[UG]{Arjan van der Schaft}\ead{a.j.van.der.schaft@rug.nl}
\address[CAM]{Information Engineering Division, University of Cambridge, CB2 1PZ, United Kingdom}
\address[UG]{Johann Bernoulli Institute for Mathematics and Computer Science, University of Groningen, 9700 AK, the Netherlands}
\address[SPLC]{Laboratoire des Signaux et Systèmes, CNRS-SUPELEC, Plateau du Moulon, 91192, Gif-sur-Yvette, France}


 %\maketitle
%
\begin{abstract}
In this paper, we examine the shifted passivity property of port-Hamiltonian systems. Shifted passivity accounts for the fact that in many applications the desired steady-state values of the input and output variables are nonzero, and thus one is interested in passivity with respect to the shifted signals. We consider port-Hamiltonian systems with strictly convex Hamiltonian, and derive conditions under which shifted passivity is guaranteed. 
In case the Hamiltonian is quadratic and state dependency appears in an affine manner in the dissipation and interconnection matrices, our conditions reduce to negative semidefiniteness of an appropriately constructed constant matrix.  Moreover, we elaborate on how these conditions can be extended to the case when the shifted passivity property can be enforced via output feedback, thus paving the path for controller design. Stability of forced equilibria of the system is analyzed invoking the proposed passivity conditions. The utility and relevance of the results are illustrated with their application to a $6$th order synchronous generator model as well as a controlled rigid body system.
\end{abstract}
\end{frontmatter}
\section{Introduction}
%\rom{For \SCL $\;$ this whole paragraph may be removed, but you asked me to keep your ``narrative" and this looks better---I think! Also, I believe there are far too many references.}. 
Passive systems are a class of dynamical systems in which the rate at which the energy flows into the system is not less than the increase in storage. In other words, starting from any initial condition, only a finite amount of energy can be extracted from a passive system. This, together with the invariance under negative feedback interconnection, has promoted passivity as 
%It is clear then that passivity is intimately related with the  stability properties of the system, making it 
a basic building block for control of dynamical and interconnected systems. 
Interested readers are referred to \cite{Arjan-book,ortega2013passivity,Bai2011} for a tutorial account of the applications of passivity in control theory.

%\rom{This is a bit biased, and incomplete, review of the literature. I'd suggest to simply quote the books.} \bow{Passivity has been used as a key tool for stability analysis and design of large scale systems and dynamic networks \cite{Arjan-book,ortega2013passivity,Willems:72,Sepulchre1}. The reason for this mainly lies in the intriguing relation of passivity with the physics of the system, and its invariance under negative feedback. These properties promote the physical energy of the system as the cornerstone of Lyapunov functions verifying stability of large scale interconnected passive systems. In a similar vein, passivity can be exploited as a powerful design tool \rom{what is this?} regulating the behavior of a system to a desired one \cite{Arjan-book,ortega2013passivity,Arcak2007,pogromsky1998passivity,pavlov2008incremental}.}

Passivity of state-space systems is commonly defined  as an input-output property for systems whose desired equilibrium state is the origin and the input and output variables are zero at this equilibrium \cite{Willems:72,Arjan-book,ortega2013passivity}. If several such systems are interconnected---for instance, a plant with a controller---the origin is an equilibrium point of the overall system whose stability may be assessed using the tools of passivity theory. In many applications, however, the desired equilibrium is not at the origin and the input and output variables of the system take nonzero values at steady-state. A standard procedure to describe the dynamics in these cases is to generate a so-called incremental model with inputs and outputs the deviations with respect to their value at the equilibrium. A natural question that arises is whether passivity of the original system is inherited by its incremental model, a property that we refer in this paper as {\em shifted passivity}. Following \cite{jayawardhana-passivity}, we use a shifted storage function to address this issue, see also the shaped Hamiltonian in \cite{maschke2000energy}. This shifted function is closely related to the notion of availability function used in thermodynamics \cite{alonso2001stabilization,keenan1951availability}.  {A byproduct of the construction of shifted storage functions is  a passivity property which is uniform for a range of steady-state solutions. This is particularly advantageous in flow networks, distribution, and electrical networks where loads/demands are not precisely known and are treated as constant disturbances, \cite{wen2003unifying,de2017bregman,trip2016internal,monshizadeh2017agreeing,JWAV:13,ferguson2017new}; see also \cite{Hines2011,Burger-duality2014,simpson2017equilibrium} where the term ``equilibrium-independent passivity" has been used to refer to the aforementioned uniform passivity property.
%a passivity property which is uniform over all possible steady-state} 
%\blue{Arjan: good to mention the Californians here with their terminology MEIP}
%This sifted storage function is defined as the Bregman distance of the Hamiltonian with respect to an equilibrium of the system \cite{bregman1967relaxation}. \rom{Is this needed?} \bow{ This function is closely related to the notion of availability function used in thermodynamics \cite{alonso2001stabilization,keenan1951availability}.}  
%, \rom{Is this needed?} \bow{ see also  \cite{ferguson2015disturbance,monshizadeh2017agreeing,de2017bregman,Arjan-book}}.

We study in this paper shifted passivity of port-Hamiltonian (pH) systems that, as is well-known, provide an attractive energy-based modeling framework for nonlinear physical systems \cite{van1995hamiltonian,Arjan2014PH,schaft2013}. The Hamiltonian readily serves as a storage function certifying passivity of a pH system, however, proving its shifted passivity is in general nontrivial. In \cite{jayawardhana-passivity} it is shown that pH systems with convex Hamiltonian are also shifted passive provided the input, dissipation and interconnection  matrices are all constant. Conditions for shifted passivity of pH systems with state-dependent matrices have been reported in \cite{maschke2000energy} and \cite{ferguson2015disturbance}. In the former case, quite conservative, integrability conditions, are imposed while the latter ones are too general and thus can be difficult to verify. The main contribution of the present paper is to give easily verifiable conditions---{\em i.e.}, monotonicity of a suitably defined function---to ensure shifted passivity of pH systems with strictly convex Hamiltonian and {\em state-dependent} dissipation and interconnection matrices. Similarly to  \cite{jayawardhana-passivity} our candidate storage function is the shifted Hamiltonian, which is associated with the Bregman distance of the Hamiltonian with respect to an equilibrium of the system \cite{bregman1967relaxation}.
%, see also  \cite{Arjan-book,de2017bregman,ferguson2015disturbance,monshizadeh2017agreeing}. 
%\rom{Is this needed?} \bow{ This function is closely related to the notion of availability function used in thermodynamics \cite{alonso2001stabilization,keenan1951availability}.}  

Notably, for the case of affine pH systems with quadratic Hamiltonian, our conditions reduce to negative semidefiniteness of an appropriately constructed {\em constant} matrix. The proposed conditions are exploited to certify local and global stability of {\em forced} pH systems, {\em i.e.}, under constant external inputs, see  \cite{maschke2000energy}.
%, arises in flow, distribution, and electrical networks where loads/demands are not precisely known and are typically treated as constant disturbances \cite{wen2003unifying,de2017bregman,trip2017optimal,JWAV:13,caliskan2014compositional,ferguson2017new}. 
An additional contribution of our work is that the proposed conditions provide an estimate of the excess and shortage of passivity that serves as a tool for controller design, see {\em e.g.} 
\cite{RO-motor-17}. 

The structure of the paper is as follows. The problem formulation is provided in Section \ref{s:pf}. The main results are given in Section \ref{s:main}, and are specialized to
quadratic affine pH systems in Section \ref{s:affine}. The results are illustrated with a synchronous generator and  a rigid body model in Section \ref{s:cases}.
The paper closes with conclusions in Section \ref{s:conclusion}.\\
 
\noindent {\bf Notation} All functions are assumed to be sufficiently smooth.  For mappings ${H}:\rea^n \to \rea$ and $\calc: \mathbb{R}^n \to \rea^{n}$ we denote the transposed gradient as $\nabla{H}:= \left(\frac{\partial {H}}{\partial x}\right)^\top $ and the transposed Jacobian matrix as $\nabla \calc:= \left(\frac{\partial \mathcal{C}}{\partial x}\right)^\top $. The Jacobian $(\nabla \calc(\cdot))^\top$ is simply denoted by $\nabla \calc(\cdot)^\top$. 
%For the distinguished vector $\overline x \in \mathbb{R}^n$ and the mapping $\calf: \mathbb{R}^n \to \rea^{n \times n}$, we define the constant  matrix $\overline \calf:=\calf(\overline x)$ and the constant vector $\nabhbar:=\nabla H(\overline x)$. 
An $n\times m$ matrix of zeros is denoted by $0_{n m}$. For a vector $x\in \R^n$, we denote its Euclidean norm by $\norm{x}$.
%\nmargin{reminder: check}
 %
%%%%%%%
\section{Problem Formulation}
\label{s:pf}
%%%%%%%%
%
Consider the pH system
%\begali{
\bse\label{sys}
\begin{align}
\label{x}
\dot x &= (J(x)-R(x)) \nabla H(x) + G u\\
\label{y}
y &= G^\top   \nabla H(x),
\end{align}
\ese
%\label{y}
%}
with state $x\in \R^n$, input $u\in R^m$, and output $y\in \R^m$. The constant matrix $G \in \rea^{n \times m}$ has full column rank, 
and $H: \rea^n \to \rea$ is the Hamiltonian of the system. The matrix $J$ is skew-symmetric, {\em i.e.}, $J(x)+J^\top(x) =0$, and 
\be\label{fpluft}
R(x)\geq R^*, \qquad \forall x\in \R^n
\ee
for some constant positive semidefinite matrix $R^*$. 
%Clearly, we have
%\be\label{fpluft}
%F(x)+F^\top  (x) \leq - 2 R^*, \qquad \forall x\in \R^n,
%\ee
%where the state-dependent matrix $F$ is given by $F(x):=J(x)-R(x)$.

Define the steady-state relation 
$$
\cale:=\{(x,u) \in \rea^n \times \rea^m \;|\; (J(x)-R(x)) \nabla H(x)+Gu=0\}.
$$
Fix $(\overline x,\overline u) \in \cale$ and the corresponding output $\overline y:=G^\top  \nabla H(\overline x)$. We are interested in finding conditions under which the mapping $(u-\overline u) \to (y - \overline y)$ is passive. We refer to this property as shifted passivity, which is formally defined next:

\begin{definition}\label{d:passive}
Consider the pH system \eqref{sys}. Let $(\overline x,\overline u) \in \cale$ and define $\overline y:=G^\top  \nabla H(\overline{x})$. The pH system \eqref{sys} is {\em shifted passive} if the mapping $(u-\overline u) \to (y - \overline y)$ is passive, {\em i.e.}, there exists a function $\calh:\R^n \rightarrow \R_{\geq 0}$ such that 
\be
%\label{e:inc-passive}
\dot \calh=(\nabla \calh)^\top  \, \dot x \leq (u-\overline u)^\top (y-\overline y)
\ee 
for all $(x, u)\in \R^n\times \R^m$.

\end{definition}

\begrem
 Note that shifted passivity is different from the classical incremental passivity property \cite{desoer2009feedback}. In fact, the latter is much more demanding as the word ``incremental" refers to two arbitrary input-output pairs of the system, whereas in the former only one input-output pair is arbitrary and the other one is fixed to a constant. 
\endrem

\section{Main Results}\label{s:main}

In this section, we provide our main results concerning shifted passivity, stability, and shifted feedback passivity of the pH system \eqref{sys}.

\subsection{Shifted passivity}

Here, we provide conditions under which the pH system \eqref{sys} is shifted passive in the sense of Definition \ref{d:passive}.
Towards this end, we make two assumptions:

\begin{assumption} 
\label{a:convexity}
The Hamiltonian $H$ is strictly convex.
\end{assumption}

%To state the next assumption, 
%%let $F(x):=J(x)-R(x)$. Then, the state dependent matrix $F$ verifies 
%%\be\label{fpluft}
%%F(x)+F^\top  (x) \leq - 2 R^*%, \qquad \forall x\in \R^n.
%%\ee
%we define the mapping $\calf:\rea^n \to \rea^{n \times n}$ as
%\begequ
%\label{calf}
%\calf(s):=F(\caln(s))
%\endequ
%where $\caln:\rea^n \to \rea^{n}$ is a left inverse of $\nabla H(x)$, i.e.,
%\begequ
%\label{caln}
%\caln(\nabh)=x,
%\endequ
%%{and for simplicity we have assumed that $\nabla H:\R^n \rightarrow \R^n$ is surjective.} 
%Note that the existence of $\caln(\cdot)$ is guaranteed by the strict convexity of $H(x)$, hence the composition of functions $F$ and $\caln$, namely $\calf$, is well-defined. In fact, $\calf$ can be constructed from $\nabla H$ following the so-called  Legendre transform as illustrated next.

%%%%%%%%%%%%%%%%%%%% Some info regarding Legendre transforms
Given the strictly convex function $H$ we define the {\it Legendre transform}, sometimes called Legendre-Fenchel transform, of $H$ as the function
\[
H^*(p) := \max_x \{x^\top p -H(x) \} ,
\]
where the domain of $H^*$ is the set of all $p$ for which the expression is well-defined ({\em i.e.}, the maximum is attained). 
We list the following properties of the Legendre transform $H^*$; see {\em e.g.} \cite{Arnold2013}, \cite{Nielsen2010}. 
\begin{enumerate}
\item
The domain of $H^*$ is equal to the convex range of $\nabla H$. 
\item
$H^*$ is strictly convex.
\item
$H^{**} = H$.
\item
$\nabla H^*( \nabla H(x))=x, \; \mbox{for all } x$.
\item $\nabla H(\nabla H^*(p))=p$, \; for all $p$ in the convex range of $\nabla H$.
\end{enumerate}
%The last property implies that in fact $\caln (s)$ is given by $\caln (s)= \nabla H^*(s)$.

Let $F(x):=J(x)-R(x)$. Leveraging the Legendre transform above, the function $F(x)$ can be restated in terms of co-energy variables $s:=\nabla H(x)$ as
\be\label{calf}
F(x)= F(\nabla H^*(s))=:\calf(s).
\ee
%%%%%%%%%%%%%%%%%%%%%%%%%%%%%%%%%%%%%%%%%%%%%%%%%%%%%%%%%%%%%%%
We denote the domain of $H^*$, which is equal to the range of $\nabla H$, by $S$.
Let $\overline s:=\nabla H(\overline x)$. We impose the following assumption on $\calf$:
\begin{assumption}
\label{a:monotonicity}
 The mapping $\calf$ verifies
\be
\label{moncon}
%\nabla (\calf(s)\,\overline s\,)+ \big(\nabla(\calf(s)\,\overline s \,)\big)^\top  - 2 R^*\leq 0,   \qquad { \forall s\in S.}
\nabla (\calf(s)\,\overline s\,)+ \nabla(\calf(s)\,\overline s \,)^\top  - 2 R^*\leq 0,   \qquad { \forall s\in S.}
\ee
\end{assumption}

Note that the choice of $R^*$ is important in feasibility of \eqref{moncon}, and is best to choose the lower bound in \eqref{fpluft} as tight as possible. Now, we have the following result:
\begin{proposition}
\label{t:passivity}
Let Assumptions \ref{a:convexity} and  \ref{a:monotonicity} hold. 
Then,  the pH system \eqref{sys} is shifted passive, namely
\be\label{e:inc-passive}
\dot \calh \leq (u-\overline u)^\top (y-\overline y)
\ee
is satisfied with
\be\label{e:shifted}
\calh(x):=H(x)-(x - \overline x)^\top   \nabla H(\overline x) - H(\overline x).
\ee
\end{proposition}
 
\BP 
First, note that $\calh$ is nonnegative as the Hamiltonian $H$ is (strictly) convex \cite{bregman1967relaxation,jayawardhana-passivity}.
%From \eqref{calf} and \eqref{caln} we have that
%\begali{
%\label{calfnabh}
%\calf(s):=F(x).
%}
Substituting \eqref{calf} into \eqref{sys} yields
$$
\dot x = \calf(s)s- \calf(\sbar)\,\sbar +  G (u-\overline u),
$$
%$$
%\dot x = (\calf(s)-R) \nabla H(x)- (\overline \calf-R)\,sbar +  G (u-\overline u),
%$$
where we have subtracted 
$
0=  \calf(\sbar)\,\sbar+ G\, \overline u.
$
%$
%0=(\overline \calf-R) \,sbar + G \overline u.
%$s
Noting that $\nabla \calh(x)= \nabla H(x) - \nabla H(\overline x)$,  the time derivative of $\calh(x)$ is computed as
%\begalis{\calf(\sbar)
\begin{align}\label{e:main-calc}
\nonumber
\dot \calh = (\nabla \calh)^\top \dot x=&\big(s-\sbar \,\big)^\top  \big(\calf(s) s- \calf(\sbar)\,\sbar\,\big)+(y-\overline y)^\top  (u-\overline u)\\
\nonumber
%\big( \calf(s) \nabla H(x)-\overline \calf\,sbar \,\big) +(y-\overline y)^\top  (u-\overline u)\\
 =& \big(s-\sbar \,\big)^\top  \big( \calf(s) - \calf(\sbar)\,\big) \sbar \\
 \nonumber& +\big(s-\sbar \,\big)^\top   F(x)\; \big(s- \sbar \,\big) +(y-\overline y)^\top  (u-\overline u)\\
\nonumber \leq&\big(s-\sbar \,\big)^\top  \big( \calf(s) - \calf(\sbar)\,\big) \sbar \\
 & - \big(s-\sbar \,\big)^\top   R^*\; \big(s- \sbar \,\big) +(y-\overline y)^\top  (u-\overline u),
 %= & \big(\nabh-\nabhbar \,\big)^\top  \big( \calf(\nabh) - \overline \calf \,\big) \nabhbar 
  %+(y-\overline y)^\top  (u-\overline u)
%\begalis{
%\dot S = &[\nabh-\nabhbar]^\top  [ \calf(\nabh) \nabla H(x)-\overline \calf\,\nabhbar] +(y-\overline y)^\top  (u-\overline u)\\
% =& [\nabh-\nabhbar]^\top  [ \calf(\nabh)\nabhbar-\overline \calf\nabhbar] \\
% & + [\nabh-\nabhbar]^\top   F(x)[\nabla H(x)- \nabhbar] +(y-\overline y)^\top  (u-\overline u)\\
% \leq & [\nabh-\nabhbar]^\top  [ \calf(\nabh)\nabhbar-\overline \calf\nabhbar] +(y-\overline y)^\top  (u-\overline u)
%}
\end{align}
%Noting that $\nabla \calh(x)= \nabla H(x) - \nabla H(x)$,  the time derivative of $\calh(x)$ is computed as
%%\begalis{
%\begin{align}\label{e:main-calc}
%\nonumber
%\dot \calh = (\nabla \calh)^\top \dot x=&\big(\nabh-\nabhbar \,\big)^\top  \big(\calf(\nabh) \nabla H(x)- \overline \calf\,\nabhbar\,\big)+(y-\overline y)^\top  (u-\overline u)\\
%\nonumber
%%\big( \calf(\nabh) \nabla H(x)-\overline \calf\,\nabhbar \,\big) +(y-\overline y)^\top  (u-\overline u)\\
% =& \big(\nabh-\nabhbar \,\big)^\top  \big( \calf(\nabh) - \overline \calf \,\big) \nabhbar \\
% \nonumber& +\big(\nabh-\nabhbar \,\big)^\top   F(x)\; \big(\nabla H(x)- \nabhbar \,\big) +(y-\overline y)^\top  (u-\overline u)\\
%\nonumber \leq&\big(\nabh-\nabhbar \,\big)^\top  \big( \calf(\nabh) - \overline \calf \,\big) \nabhbar \\
% & - \big(\nabh-\nabhbar \,\big)^\top   R^*\; \big(\nabla H(x)- \nabhbar \,\big) +(y-\overline y)^\top  (u-\overline u),
% %= & \big(\nabh-\nabhbar \,\big)^\top  \big( \calf(\nabh) - \overline \calf \,\big) \nabhbar 
%  %+(y-\overline y)^\top  (u-\overline u)
%%\begalis{
%%\dot S = &[\nabh-\nabhbar]^\top  [ \calf(\nabh) \nabla H(x)-\overline \calf\,\nabhbar] +(y-\overline y)^\top  (u-\overline u)\\
%% =& [\nabh-\nabhbar]^\top  [ \calf(\nabh)\nabhbar-\overline \calf\nabhbar] \\
%% & + [\nabh-\nabhbar]^\top   F(x)[\nabla H(x)- \nabhbar] +(y-\overline y)^\top  (u-\overline u)\\
%% \leq & [\nabh-\nabhbar]^\top  [ \calf(\nabh)\nabhbar-\overline \calf\nabhbar] +(y-\overline y)^\top  (u-\overline u)
%%}
%\end{align}
where we used \eqref{y} in the first identity, added and subtracted the term  $\big(s-\sbar\,\big)^\top \big(\calf(s) \sbar\,\big)$ and used \eqref{calf}  to write the second equality, while the bound is obtained invoking \eqref{fpluft}.
Now, let %$\calm:\R^n \rightarrow \R^n$ be defined as
\be\label{e:calm}
\calm(s):=\calf(s) \sbar - R^*s.
\ee
Then $\dot \calh$ can be written as
\be\label{e:dissipation}
\dot \calh= \big(s-\sbar \,\big)^\top  \big( \calm( s) - \calm(\sbar) \big)+(y-\overline y)^\top  (u-\overline u).
\ee
By \eqref{moncon}, we have that $\nabla \calm(s) + (\nabla \calm(s))^\top \leq 0$, for all {$s\in S$}, which ensures that the map $ \calm(\cdot)$ is monotone \cite{ryu2016primer}. The proof is completed noting that by monotonicity
 $$
\big(s-\sbar \,\big)^\top  \big( \calm( s) - \calm(\sbar) \big)\leq 0.
 $$
 \EP

%\vspace{-2mm}
\begrem
By Assumptions \ref{a:convexity} and \ref{a:monotonicity}, both the strict convexity and the monotonicity property must hold for the whole sets $\R^n$ and $S$, respectively, which results in ``global" shifted passivity of \eqref{sys}. For local shifted passivity,\footnote{By ``local" we mean that there exist open neighborhoods $\calx\subseteq \R^n$ and $\calu\subseteq \R^m$ of $(\overline x, \overline u)\in \calx \times \calu$ such that \eqref{e:inc-passive} holds for all $(x, u)\in \calx \times \calu$.} we can restrict to a subset $\calx \subseteq \R^n$, with Assumptions \ref{a:convexity} and \ref{a:monotonicity} modified to
\begin{enumerate}
\item The Hamiltonian is strictly convex in $\calx\subseteq \R^n$.
\item  Inequality \eqref{moncon} holds for all $s\in \cals:=\{\nabla H(x) \mid x\in \calx\},$
\end{enumerate}
while $R^*$ is any matrix satisfying, instead of \eqref{fpluft}, $R(x)\geq R^*, \quad \forall x\in \calx$. 
\endrem

%it suffices that $\nabla^2 H (\overline x)>0$ and \eqref{moncon} holds in an open neighborhood $\cals\subseteq \R^n$ of $\sbar$.  Consistently, the dissipation matrix is \red{assumed} \p{I think we should rephrase this, since it is not our assumption, but the structure of the system.} to satisfy $R(x)\geq R^*$ for all $x\in \calx=\{{x} \mid \nabla H({x})\in \cals\}$, which means that \eqref{fpluft} holds for all $x\in \calx$.

%\footnote{The set $\cals$ needs to be convex to ensure that $\nabla H$ is injective, and the map $\calf$ is well-defined.}
%\footnote{The set $\cals$ needs to be convex to ensure that  the associated Legendre transformation is injective, and thus the map $\calf$ is well-defined. %For nonconvex sets, additional conditions are needed \cite{}.} 
%\be\label{e:monotone-sbar}
%\nabla \overline \calk+ (\nabla \overline \calk\,)^\top <0, \qquad \calk(s):=\calf(s)\nabhbar, \quad \overline s= \nabhbar.
%%\big(\nabla (\overline{\calf(s)\nabhbar}\,)\big)^\top  < 0,   \qquad s=\nabhbar.
%\ee 
%%\be\label{e:monotone-sbar}
%%\nabla (\overline{\calf(s)\nabhbar}\,)+\big(\nabla (\overline{\calf(s)\nabhbar}\,)\big)^\top  < 0,   \qquad s=\nabhbar.
%%\ee 

\subsection{Stability of the forced equilibria}
%\red{Romeo: This is not clear. Do you refer to the system with u =hat u? This paragraph should be rewritten.}
Lyapunov stability of the equilibrium of \eqref{sys} with $u=\overline u$, immediately follows from Proposition \ref{t:passivity}, with the Lyapunov function being the shifted Hamiltonian $\calh$. Moreover, asymptotic stability follows by imposing the condition that $\dot\calh$ is negative definite.
Below, we provide the results concerning stability of the forced pH-system \eqref{x} with $u=\overline u$. 
Although deducing stability properties from passivity is well-known \cite{Arjan-book}, we provide the proof for the sake of completeness.
%For asymptotic stability, however, the previous conditions need to be suitably modified. This is discussed in the following theorem.
%Deducing stability properties of the system from passivity is well-known
%Next, we provide the results concerning stability of the forced pH-system \eqref{x} with $u=\overline u$. 
%First, note that Lyapunov stability of the equilibrium immediately follows from Theorem \ref{t:passivity}, with the Lyapunov function being the shifted Hamiltonian $\calh$. 
%For asymptotic stability, however, the previous conditions need to be suitably modified. This is discussed in the following theorem.

%\rom{This theorem and its ``proof" are totally unnecessary. It is VERY well known that the equilibrium is asymptotically stable if $\dot \calh$ is negative definite and this is global if $\calh$ is proper. That's all we need to say. Also, this notation of $\bar s$ is a mess.} 
{\begin{proposition}\label{t:stability}
Consider the pH system \eqref{x} for some constant input $u=\overline u$, and let $(\overline x, \overline u)\in \cale$.
% Let Assumptions \ref{a:3} and \ref{a:4} hold. Then the equilibrium $\overline x$, $(\overline x, \overline u)\in \cale$, is globally asymptotically stable.
%Let $\nabla^2 H(\overline x)>0$, $(\overline x, \overline u)\in \cale$. 
Then, we have
\begin{enumerate}
%\item The equilibrium is Lyapunov stable if $\nabla^2 H(\overline x)>0$ and inequality \eqref{moncon} holds for all $s\in \cals$, with $\cals\subseteq \R^n$ being an open neighborhood of $\nabhbar$. 
\item The equilibrium is asymptotically stable if $\nabla^2 H(\overline x)>0$ and there exists $\epsilon>0$ such that the inequality
\be\label{e:s-moncon}
\nabla (\calf(s)\,\overline s\,)+\nabla (\calf(s)\, \overline s\,)^\top  -2 R^*\leq -2\epsilon I_n,  
\ee
holds at $s= \overline s$.\footnote{This means that the Jacobian of $\calf(s)\,\overline s\,$ in \eqref{e:s-moncon} has to be evaluated at $s=\overline s$.}
%Assumption \ref{a:s-monotone} holds with $\cals=\{\nabla \overline H\}$.
\item The equilibrium is globally asymptotically stable if the Hamiltonian $H$ is strongly convex and \eqref{e:s-moncon} holds for all $s\in \R^n$.
\end{enumerate}
\end{proposition}

\BP
Set $u=\overline u$ and take the shifted Hamiltonian $\calh$ as the Lyapunov candidate, and suppose that \eqref{e:s-moncon} holds at the point $s= \overline s$. Then there exists an open neighborhood 
$\cals$ of $\overline s$ and some $0<\epsilon'\leq \epsilon$ such that 
\be\label{e:s-moncon2}
\nabla (\calf(s)\overline s\,)+\nabla (\calf(s)\overline s \,)^\top -2R^*\leq -2\epsilon' I_n,  
\ee
for all $s \in \cals$. Let $\calx:=\{x\mid \nabla H(x)\in \cals\}.$ By $\nabla H^*(\overline s)=\overline x$,  and continuity of $\nabla H^*$\footnote{The map $\nabla H^*$ is still well-defined as $H$ is locally strictly convex, {\em i.e.}, $\nabla^2 H(\overline x)>0$, {and $\cals$ can be chosen as a convex set.}}, the set $\calx\in \R^n$ defines an open neighborhood of the equilibrium $\overline x$.
It is easy to see that  \eqref{e:s-moncon2} implies (local) strong monotonicity of the map $\calm$ in \eqref{e:calm}, see \cite{ryu2016primer}, and the dissipation inequality \eqref{e:dissipation} gives 
%Therefore, by Theorem \ref{t:passivity} and equality \eqref{e:dissipation}, we find that 
\[
\dot \calh= -\epsilon' \norm{s-\sbar}^2, \qquad \forall x\in \calx.
\]
%\red{Romeo!: define eucledean norm}
Now, noting that $H$ is locally strictly convex, the function $S$ is locally nonnegative and is equal to zero whenever $x=\overline x$ \cite{bregman1967relaxation}. Hence, there exists a compact subset of $\calx$ which is forward invariant along the solutions of the system (see also \cite[Prop. 2]{jayawardhana-passivity}). By invoking LaSalle's invariance principle, on the invariant set we have $s=\sbar$, which results in $x=\overline x$ by strict convexity of $H$.

To prove the second statement, it suffices to show that $S$ is radially unbounded.  This follows from strong convexity of $H$ noting that \cite[Ch.2]{nesterov2013introductory}
\[
\calh(x)= H(x)-(x - \overline x)^\top   \nabla H(\overline x) - H(\overline x) \geq \mu \norm{x-\overline x}^2,
\]
for some $\mu\in\R^+$.
\EP}

\begrem
The identity matrix in the right hand side of \eqref{e:s-moncon} can be replaced by a positive semidefinite matrix $C^\top  C$, with $C\in \R^{m\times n}$, if the equilibrium is ``observable" from the input-output pair $(\overline u, C\nabla H(\overline x))$, namely  if
\[
\dot x= F(x)\nabla H+g \overline u, \;\; C \nabh = C\nabla H(\overline x) \;\;\Longrightarrow \;\;x=\overline x.
\]
\endrem

\subsection{Enforcing shifted passivity via output feedback}

We complete this section by considering the case where the condition \eqref{moncon} does not hold, which means that the system \eqref{sys} may not be shifted passive, but it can be rendered shifted passive via output feedback. 
%\rom{OUT: In this case, one can still quantify the possible shortage of passivity using a condition similar to \eqref{moncon}.  
%This results in a shifted {\em feedback passivity} property as stated in the following proposition.}

\begin{proposition}\label{p:dissipate}
Consider the pH system \eqref{sys} verifying Assumption \ref{a:convexity} and such that 
\be
\nonumber
\nabla (\calf(s)\,\overline s\,)+\nabla (\calf(s)\, \overline s\,)^\top  -2 R^*\leq 2\gamma\, GG^\top, 
\ee
for some $\gamma \in \R$.
Then, the shifted Hamiltonian \eqref{e:shifted} satisfies the following dissipation inequality
\be
\nonumber
\dot \calh \leq (u-\overline u)^\top (y-\overline y) + \gamma \norm{y-\overline y}^2.
\ee
\end{proposition}

\BP
The proof is analogous to that of Proposition \ref{t:passivity}, by adding and subtracting the term $\gamma GG^\top (s- \sbar )$ in \eqref{e:main-calc}, and modifying the map $\calm$ as 
\[
 \widetilde{\calm}(s):=\calm(s)- \gamma  GG^\top s.
\]
\EP
\vspace{-2mm}
\begrem
Note that a negative $\gamma$ proves that the pH system is (output-strictly) shifted passive. On the other hand, a positive $\gamma$ indicates the shortage of shifted passivity. Notice that the simple proportional controller
$$
u  =\overline u - K_P (y-\overline y)  \,+\, v,
$$
with $K_P \geq \gamma I$, ensures that the interconnected system is passive from the external input $v$ to output $y-\overline y$. {Analogously, Proposition 7 can be used to design dynamic passive controllers to stabilize} the closed-loop system, see \cite{RO-motor-17} for an application to control of permanent magnet synchronous motors.
%in \rom{PMSM paper} to prove that PI control of permanent magnet synchronous motors is globally stabilizing.  
\endrem
%
%
%%%%%%%
\section{Application to Quadratic Affine Systems\label{s:affine}}
In this section we specialize our results to the case where 
\begequ
\label{fx}
F(x)=F_0+ \sum_{i=1}^n F_i x_i,
\endequ
with $F_j \in \rea^{n \times n},\;j=0,\dots,n,$ constant
% and verifying $F_j=-F_j^\top $ 
and 
\begequ
\label{xqx}
H(x)=\hal x^\top   Q x,
\endequ
with $Q \in \rea^{n \times n}$ being {positive definite}. We call these systems quadratic affine pH systems.
%\rom{Not clear at all. This restriction of constant $R$ is because we are doing a wrong bounding. Remark 12 seems to indicate this.} 

In order to satisfy \eqref{fpluft} and state the global version of our results, we need to assume that $R(x)=R_0$ for some constant matrix $R_0$. 
 This is due to the fact that in the affine case the inequality $R(x)\geq R^*$, for all $x\in \R^n$, implies that the matrix $R$ is constant.
In Remark \ref{r:Rx}, we elaborate on how this assumption is relaxed to obtain local results. Note that, in this case, $F_0+F_0^\top = -2R_0 \leq 0$ and $F_j+F_j^\top=0$ for each $j\geq1$.

\begin{proposition}\label{p:quad-affine}
	Consider the quadratic affine pH system \eqref{sys} with \eqref{fx} and \eqref{xqx}.  Fix $(\overline x,\overline u) \in \cale$ and define the $n \times n$ constant matrix
	\begali{
		\label{e:b}
		B:=\sum_{i=1}^n  F_i Q \, \overline x \, e^\top_iQ^{-1},
	}
	with $e_i \in \rea$ the $i$-th element of the orthogonal basis. If
	\begequ
	\label{c:quadraticaffine}
	B+B^\top  -2 R_0 \leq 0,
	\endequ
	then  \eqref{sys} is shifted passive, namely
	$$
	\dot\calh  \leq (y-\overline y)^\top  (u-\overline u),
	$$
	where $\calh$ is the quadratic shifted Hamiltonian function 
	$$
	\calh(x):=\hal (x - \overline x)^\top   Q (x - \overline x).
	$$
\end{proposition}

\BP The proof follows by verifying the conditions of Proposition \ref{t:passivity}. In this case, {$H^*(p) = \frac{1}{2}p^\top Q^{-1}p$,}
	 \begalis{
		\nabla H^*(s) =Q^{-1}s, \qquad s=Qx,
	}
and
	\begali{\label{e:F(s)}
		\calf(s) =F_0+ \sum_{i=1}^n F_i (e^\top  _iQ^{-1}s). 
	}
	Hence,
	%\begalis{
	\[
		\calf(s)\sbar  =(F_0+ \sum_{i=1}^n F_i (e^\top  _iQ^{-1}s))Q\,\overline x.
	\]
	%}
	Now, by rewriting the last expression in the equivalent form
	\begalis{
		\calf(s)\sbar  =F_0Q\overline x+ \sum_{i=1}^n F_iQ\overline x e^\top  _iQ^{-1}s,
	}
	we obtain that
	$$
	\nabla (\calf(s)\sbar)=\sum_{i=1}^n F_iQ\overline x e^\top  _iQ^{-1}.
	$$
	Finally, using \eqref{e:b}, condition  \eqref{moncon} takes the form
	\begalis{
		\nabla (\calf(s)\sbar)+\nabla (\calf(s)\sbar)^\top  -2R_0
		%& = \sum_{i=1}^n F_iQ\overline x e^\top  _iQ^{-1} +Q^{-1}[\sum_{i=1}^n F_iQ\overline x e^\top  _i]^\top   -2R\\
		& = B+B^\top  -2 R_0 \leq 0.
	}
\EP
\begrem\label{r:ins}
	The stability condition in Proposition \ref{p:quad-affine} can equivalently be stated in terms of the co-energy variables 
	$s=\nabla H(x)$, which in certain cases decreases the computational effort.  
	To this end, note that by \eqref{e:F(s)}, the function $\calf(s)$ can be written in the affine form:
	%\begequ
	%\label{e:fx}
	\[ \calf(s)=F_0+ \sum_{i=1}^n \calf_i s_i,\]
	where $\calf_i:=\sum_{j=1}^{n} F_j Q^{-1}_{ij}$.
	Hence, the matrix  $B$ in \eqref{e:b} can be equivalently written as
	\[
	%\label{e:bF}
	B:=\sum_{i=1}^n  \calf_i\,Q\overline x \,e^\top  _i.
	\]
\endrem

Noting that $H$ is strongly convex and the condition \eqref{c:quadraticaffine} is state independent, Proposition \ref{t:stability} yields the following result:
\begin{corollary}\label{c:quadraticaffinestability}
Consider the quadratic affine pH system \eqref{x} with \eqref{fx} and \eqref{xqx} under some constant input $u=\overline u$, and let $(\overline x, \overline u)\in \cale$.
The equilibrium $\bar x$ is globally asymptotically stable if \eqref{c:quadraticaffine} holds with strict inequality. 
\end{corollary}
{
\begin{remark}\label{r:Rx}
Analogous to the previous section, local variations of Proposition \ref{p:quad-affine} and Corollary \ref{c:quadraticaffinestability} can be obtained by restricting $x$ in a domain $\calx\in \R^n$ with $\overline x \in \calx$. In that case, the matrix $R_0$ in \eqref{c:quadraticaffine} is replaced by $R^*$, where $R(x)\geq R^*\geq 0$ for all $x\in \calx$. 
\end{remark}}

\section{Case Studies}\label{s:cases}
In this section, we apply the proposed method to two physical systems. Both systems are affine and have quadratic Hamiltonian.
\subsection{Synchronous generator ($6^{th}$-order model) connected to a resistor}
The state variables of the six-dimensional model of the synchronous generator comprise of the stator fluxes on the $dq$ axes $\psi_d \in \R,\psi_q \in \R$, rotor fluxes $\psi_r\in \R^3$ (the first component of $\psi_r$ corresponds to the field winding and the remaining two to the damper windings), and the angular momentum of the rotor $p$. The Hamiltonian H (total stored energy of the synchronous generator) is the sum of the magnetic energy of the generator and the kinetic energy of the rotating rotor. More precisely, the Hamiltonian takes the form $H(x)=\half x^\top Qx$ with $x=\begin{bmatrix}
{\psi_d}& {\psi_q}&{\psi_r}& {p}
\end{bmatrix}^\top$ and
\begin{align*}
Q=\begin{bmatrix}
L^{-1} & 0_{51}
\\
0_{15} & m^{-1}
\end{bmatrix}>0
\;,\;\;
L&=\begin{bmatrix}
L_d        &       0           &         kL_{afd}        &      kL_{akd}      &     0
\\
0         &      L_q           &        0              &     0            &   -kL_{akq}
\\
kL_{afd}    &      0            &        L_{ffd  }         &     L_{akd}         &    0
\\
kL_{akd}    &      0             &       L_{akd}           &     L_{kkd}         &    0
\\
0         &      -kL_{akq}        &      0             &      0           &     L_{kkq}
\end{bmatrix}\;,
\end{align*}
where $m \in \R$ is the total moment of inertia of the turbine and the rotor. Note that the elements of the inductance matrix $L$ are all constant parameters,  see \cite{Shaik2013} for more details. The system dynamics is then given by the pH system \cite{Shaik2013}
\begin{align*} 
\dot x=
%\begin{bmatrix}
%\dot{\psi_d}\\ \dot{\psi_q}\\ \\ \dot{\psi_r}\\ \\ \dot{p}
%\end{bmatrix}
%=
(J(x)-R) \nabla H(x)
%\begin{bmatrix}
%\rond{H}{\psi_d}\\ \rond{H}{\psi_q} \\ \\ \rond{H}{\psi_r}\\ \\ \rond{H}{p}
%\end{bmatrix}
+
G
\begin{bmatrix}
V_f\\
\tau
\end{bmatrix}\;,
\end{align*}
with
\begin{align*}
J(x)&=\begin{bmatrix}
0_{22}
&
0_{23}
&
\begin{bmatrix}
-\psi_q
\\
\psi_d
\end{bmatrix}
\\ \\
0_{32} & 0_{33} & 0_{31}
\\ \\ 
\begin{bmatrix}
\psi_q
&
-\psi_d
\end{bmatrix}
& 0_{13} & 0
\end{bmatrix}
,\;
R=
\begin{bmatrix}
\begin{bmatrix}
r & 0
\\
0 & r
\end{bmatrix}
&
0_{23}
&
0_{21}
\\ \\
0_{32} & \begin{bmatrix}
R_f & 0 & 0
\\
0 & R_{kd} & 0
\\
0 & 0 & R_{kq}
\end{bmatrix} & 0_{31}
\\ \\ 
0_{12}
& 0_{13} & d
\end{bmatrix}>0\;
,
\quad
G=\begin{bmatrix}
0_{21} & 0_{21}
\\
\begin{bmatrix}
1\\0\\0
\end{bmatrix} & 0_{31}
\\
0 & 1
\end{bmatrix}\;,
\end{align*}
where $V_f$ represents the rotor field winding voltage, $\tau$ is the mechanical torque, $r$ is the summation of the load and stator resistances, $R_f,R_{kd},R_{kq}$ denote the rotor resistances, and $d$ corresponds to the mechanical friction. 
We can rewrite the system as
\begin{align}\label{e:sync-gen-s}
Q^{-1} \dot s
=(\calj(s)-R)s+
G
\begin{bmatrix}
V_f\\
\tau
\end{bmatrix}\, ,
\end{align}
where $s=Qx=[I_d\;\; I_q\;\;I_r\;\;\w]^\top $. Here $I_d\in \R,I_q\in \R$ are the components of the stator current on the $dq$ axes, and $I_r\in \R^3$ and $ \w \in \R$ are the currents and angular velocity of the rotor, respectively.
Note that 
\begin{align*}
\calj(s)&=
\begin{bmatrix}
0_{22}
&
0_{23}
&
v_J(s)
\\ \\
0_{32} & 0_{33} & 0_{31}
\\ \\ 
v^\top _J(s)
& 0_{13} & 0
\end{bmatrix},\qquad
v_J(s):=\begin{bmatrix}
-L_qI_q+L_{akq}I_{kq}
\\[1mm]
L_dI_d + L_{afd}I_f + L_{akd} I_{kd}
\end{bmatrix}\;\; .
\end{align*}
Let $V_f=\overline V_f$ and $\tau=\overline \tau$, for some constant vectors $\overline V_f$ and $\overline \tau$. 
Through straightforward calculations, and using Remark \ref{r:ins}, the condition \eqref{c:quadraticaffine} reads as
%\begin{equation}\label{e:condition2}
%\resizebox{8.7cm}{!}{$
%\begin{bmatrix}
%-2r & \overline \w (L_d-L_q) & 0 & 0 & k\overline \w L_{akq} & - \bar I_q L_d 
%\\
%\overline \w (L_d-L_q) & -2r & k\overline \w L_{afd} & k\overline \w L_{akd} & 0 &  \bar I_d L_d 
%\\
%0  & k\overline \w L_{afd} & -2R_f & 0 & 0&  - \bar I_q L_{afd} 
%\\
%0  & k\overline \w L_{akd} & 0& -2R_{kd} &  0&  - \bar I_q L_{akd} 
%\\
%k\overline \w L_{akq} & 0& 0 &  0& -2R_{kq} & - \bar I_d L_{akq} 
%\\
%- \bar I_q L_d  & \bar I_d L_d & - \bar I_q L_{afd}  &  - \bar I_q L_{akd}& - \bar I_d L_{akq}  & - 2d
%\end{bmatrix} < 0.$}
%\end{equation}
\begin{equation}\label{e:condition2}
\begin{bmatrix}
-2r & \overline \w (L_d-L_q) & 0 & 0 & k\overline \w L_{akq} & - \bar I_q L_d 
\\
\overline \w (L_d-L_q) & -2r & k\overline \w L_{afd} & k\overline \w L_{akd} & 0 &  \bar I_d L_d 
\\
0  & k\overline \w L_{afd} & -2R_f & 0 & 0&  - \bar I_q L_{afd} 
\\
0  & k\overline \w L_{akd} & 0& -2R_{kd} &  0&  - \bar I_q L_{akd} 
\\
k\overline \w L_{akq} & 0& 0 &  0& -2R_{kq} & - \bar I_d L_{akq} 
\\
- \bar I_q L_d  & \bar I_d L_d & - \bar I_q L_{afd}  &  - \bar I_q L_{akd}& - \bar I_d L_{akq}  & - 2d
\end{bmatrix} \leq 0\;,
\end{equation}
where $\bar I_d, \bar I_q,\bar I_r,\overline \w$ are the associated values of $s$ at the equilibrium of \eqref{e:sync-gen-s}, {\em i.e.} 
\begin{align*}
(\calj(\overline s)-R)\overline  s+
G
\begin{bmatrix}
\overline V_f\\
\overline \tau
\end{bmatrix}=0\,, 
\end{align*}
with $\overline s=[\bar I_d\;\; \bar I_q\;\;\bar I_r\;\;\overline \w]^\top $.
Hence, by Proposition \ref{p:quad-affine}, \eqref{e:sync-gen-s} is shifted passive if \eqref{e:condition2} holds. Moreover, by Corollary \ref{c:quadraticaffinestability}, if \eqref{e:condition2} holds with strict inequality, then the equilibrium $\overline x= Q^{-1} \overline s$ is globally asymptotically stable. The stability result is consistent with those of \cite{caliskan2014compositional,Arjan2016}\footnote{Notice that there is a typo in \cite{Arjan2016} as the term $\overline \w (L_d-L_q)$ is missing.}. Note that Corollary \ref{c:quadraticaffinestability} is valid for a general quadratic affine pH-system, and the condition \eqref{e:condition2} is obtained in a systematic manner here, namely by verifying the negative definiteness test in \eqref{c:quadraticaffine}. Moreover, if \eqref{e:condition2} does not hold, then in view of Proposition \ref{p:dissipate}, one can investigate the possibility of designing suitable proportional, PI, or more generally dynamic (input-strictly) passive controllers rendering the equilibrium globally asymptotically stable.

%{\tcb If \eqref{e:condition2} does not hold, then in view of Proposition \ref{p:dissipate}, there exists a proportional controller 
%(with sufficiently high gain) rendering the equilibrium globally asymptotically stable if 
%\[
%\begin{bmatrix}
%-2r & \overline \w (L_d-L_q) & 0 & k\overline \w L_{akq} 
%\\
%\overline \w (L_d-L_q) & -2r & k\overline \w L_{akd} & 0 
%\\
%%0  & k\overline \w L_{afd} & 0 & 0&  - \bar I_q L_{afd} 
%%\\
%0  & k\overline \w L_{akd} & -2R_{kd} &  0
%\\
%k\overline \w L_{akq} & 0 &  0& -2R_{kq} 
%%\\
%%- \bar I_q L_d  & \bar I_d L_d & - \bar I_q L_{afd}  &  - \bar I_q L_{akd}& - \bar I_d L_{akq}  & - 2d
%\end{bmatrix} < 0\;,
%\]
%which reduces to %negative semidefinitess of the $2\times2$ matrix
%\[
%\begin{bmatrix}
%-2r+\frac{(k\overline \w L_{akq})^2}{2R_{kq}} & \overline \w  (L_d-L_q)\\
%\overline \w (L_d-L_q) & -2r+\frac{(k\overline \w L_{akd})^2}{2R_{kd}}
%\end{bmatrix}<0. 
%\]
%}
%\red{Arjan: express clearly what is the contribution of Section 5.1 wrt [29, 30] e.g. shifted passivity and systematic way to obtain the condition.}
%The problem below is of interest:
%\begin{problem}
%Find $c$ such that
%\[
%\norm{z}^2\leq c \rightarrow z \in \mathcal{D},
%\]
%where $\mathcal{D}$ is the domain of monotonicity characterized in Theorem \ref{t:main}.
%\end{problem}
\subsection{Controlled rigid body under constant disturbances}
\begin{figure}
	\centering
	\includegraphics[width=3cm]{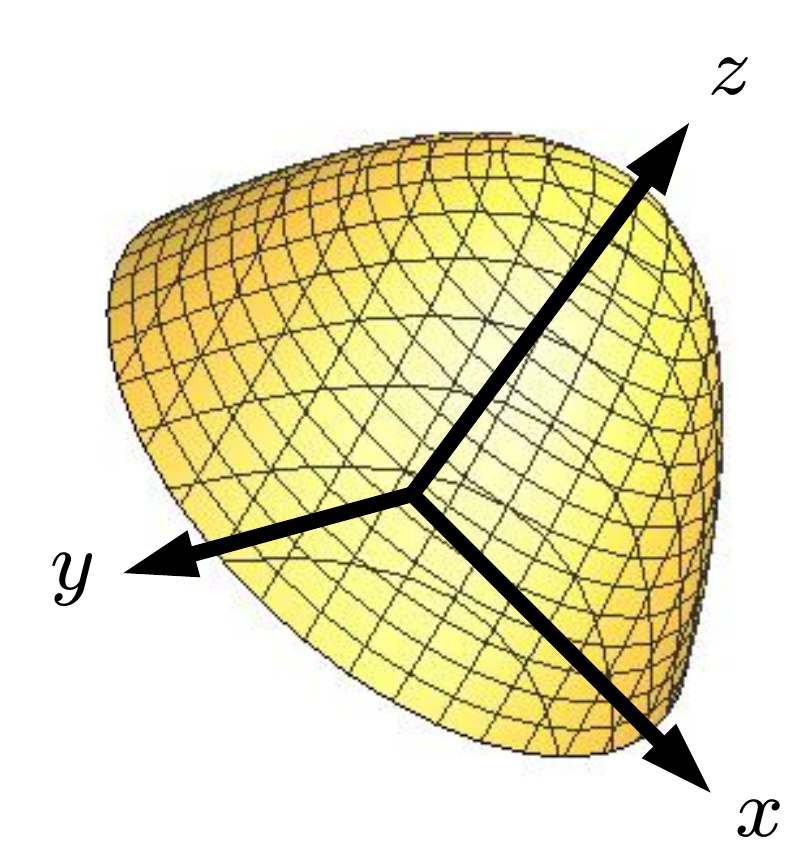}
	\caption{A rigid body with three axes of rotation}
	\label{f:rb}
\end{figure}
The equations for the angular momentum of a rigid body (see Figure \ref{f:rb}) with external torque $u \in \R^{3}$ and disturbance $d \in \R^{3}$ reads as \cite{Arjan2014PH}
\begin{align}
\dot p&=
J(p)
\nabla H(p)
%\begin{bmatrix}
%\rond{H}{p_x}
%\\
%\rond{H}{p_y}
%\\
%\rond{H}{p_z}
%\end{bmatrix}
+
u+d\\
y&=\nabla H(p)\;,
\end{align}
where $p=
\begin{bmatrix}
p_x&p_y&p_z
\end{bmatrix}^\top$,  
$$
J(p)=\begin{bmatrix}
0 & -p_z & p_y
\\
p_z & 0 & -p_x
\\
-p_y & p_x & 0
\end{bmatrix}\;,$$
and the Hamiltonian is given by 
$H(p)=\half p^\top  M p$,
with 
\[
M=
\begin{bmatrix}
m_x & 0 &0
\\
0&m_y&0
\\
0&0&m_z
\end{bmatrix}>0\;.
\]
%
%\begin{align}
%\begin{bmatrix}
%p_x
%\\
%p_y
%\\
%p_z
%\end{bmatrix}
%=
%I
%\begin{bmatrix}
%\w_x
%\\
%\w_y
%\\
%\w_z
%\end{bmatrix}\;,
%\end{align}
Here, $m_x\in \R$, $m_y\in \R$, and $m_z\in \R$ are the principal moments of inertia. Consider a constant disturbance $d=\begin{bmatrix}
d_x&d_y&d_z
\end{bmatrix}^\top $ and a proportional controller $u=-Ry$ with $R=\diag\, (r_x,r_y,r_z)>0$.
We can rewrite the system as
\begin{align}\label{e:rigid}
\begin{aligned}
M\dot \w
=
\calj(\w)-R
\w
+
\begin{bmatrix}
d_x\\d_y\\d_z
\end{bmatrix}\;.
\end{aligned}
\end{align}
where 
$$
\calj(\w)=
\begin{bmatrix}
0 & -\w_zm_z & \w_ym_y
\\
\w_zm_z &0 & -\w_xm_x
\\
-\w_ym_y & \w_xm_x & 0
\end{bmatrix}\;,
$$
and $\w=M^{-1}p=\begin{bmatrix}
\w_x&\w_y&\w_z
\end{bmatrix}^\top $ is the vector of angular velocities around the axes $x$, $y$, and $z$.
The point $(\overline \w_x,\overline \w_y,\overline \w_z)$ is an equilibrium of the system \eqref{e:rigid} satisfying
\begin{align*}
\calj(\overline \w)-R\,
\overline \w
+
\begin{bmatrix}
d_x\\d_y\\d_z
\end{bmatrix}
=0\;,
\end{align*}
with $\overline \w=\begin{bmatrix}
\overline\w_x&\overline\w_y&\overline w_z
\end{bmatrix}^\top$. Through straightforward calculations, the condition \eqref{c:quadraticaffine} (with strict inequality) reads as
\begin{align}
\begin{bmatrix}\label{e:RigidCondition}
-2r_x & \overline \w_z(m_y-m_x) & \overline \w_y(m_x-m_z)
\\
\overline \w_z(m_y-m_x) & -2r_y & \overline \w_x(m_z-m_y)
\\
\overline \w_y(m_x-m_z) & \overline \w_x(m_z-m_y) & -2r_z
\end{bmatrix}< 0\;.
\end{align}
Hence, by Corollary \ref{c:quadraticaffinestability}, if \eqref{e:RigidCondition} holds for the equilibrium point $(\overline \w_x,\overline \w_y,\overline \w_z)$, then global asymptotic stability is guaranteed. In the case that there is disturbance actuating only on one axis, {\em e.g.} $d_y=d_z=0$ (without loss of generality), the equilibrium $(\w_x,\w_y,\w_z)=(\frac{d_x}{r_x},0,0)$ is globally asymptotically stable if
$$r^2_xr_yr_z> \Big(\frac{d_x(m_z-m_y)}{2}\Big)^2\;.$$

\section{Conclusion}\label{s:conclusion}
We have examined the shifted passivity property of pH systems with convex Hamiltonian by proposing conditions in terms of the monotonicity of suitably constructed functions. 
We have leveraged these conditions to study (global) asymptotic stability of forced equilibria of the system. As we observed, for quadratic affine pH system, shifted passivity and (global) asymptotic stability are guaranteed if  an appropriately constructed constant matrix is negative semidefinite.  We demonstrated the applicability and usefulness of the results on a $6$th order synchronous generator model and a controlled rigid body system. Future works include attempting to reduce possible conservatism in the stability conditions as well as investigating the connections of the proposed results to  contraction and differential passivity \cite{forni2014differential,forni2013differential}. \\
%%%%%%%%%
%%

%\rom{There're many typos in the references (e.g., Van is van) and the notation is not homogeneous: sometimes caps, others no; use of blanks, etc. Please correct.} 
%)
\section{References}
\bibliographystyle{IEEEtran}
\bibliography{nima2}

\begin{thebibliography}{10}
\providecommand{\url}[1]{#1}
\csname url@samestyle\endcsname
\providecommand{\newblock}{\relax}
\providecommand{\bibinfo}[2]{#2}
\providecommand{\BIBentrySTDinterwordspacing}{\spaceskip=0pt\relax}
\providecommand{\BIBentryALTinterwordstretchfactor}{4}
\providecommand{\BIBentryALTinterwordspacing}{\spaceskip=\fontdimen2\font plus
\BIBentryALTinterwordstretchfactor\fontdimen3\font minus
  \fontdimen4\font\relax}
\providecommand{\BIBforeignlanguage}[2]{{%
\expandafter\ifx\csname l@#1\endcsname\relax
\typeout{** WARNING: IEEEtran.bst: No hyphenation pattern has been}%
\typeout{** loaded for the language `#1'. Using the pattern for}%
\typeout{** the default language instead.}%
\else
\language=\csname l@#1\endcsname
\fi
#2}}
\providecommand{\BIBdecl}{\relax}
\BIBdecl

\bibitem{Arjan-book}
A.~van~der Schaft, \emph{L2-Gain and Passivity Techniques in Nonlinear
  Control}.\hskip 1em plus 0.5em minus 0.4em\relax 3rd Revised and Enlarged
  Edition (1st edition 1996, 2nd edition 2000), Springer Communications and
  Control Engineering series, Springer-International, 2017.

\bibitem{ortega2013passivity}
R.~Ortega, J.~A.~L. Perez, P.~J. Nicklasson, and H.~Sira-Ramirez,
  \emph{Passivity-based Control of Euler-Lagrange Systems: Mechanical,
  Electrical and Electromechanical Applications}.\hskip 1em plus 0.5em minus
  0.4em\relax Springer Science \& Business Media, 2013.

\bibitem{Bai2011}
H.~Bai, M.~Arcak, and J.~Wen, \emph{Cooperative Control Design: {A} Systematic,
  Passivity--based Approach}.\hskip 1em plus 0.5em minus 0.4em\relax New York,
  NY: Springer, 2011.

\bibitem{Willems:72}
J.~Willems, ``{Dissipative dynamical systems Part I : General Theory},''
  \emph{{Archive for Rational Mechanics and Analysis}}, vol.~{45}, pp.
  {321--351}, {1972}.

\bibitem{jayawardhana-passivity}
B.~Jayawardhana, R.~Ortega, E.~Garc{\'\i}a-Canseco, and F.~Casta{\~{n}}os,
  ``Passivity of nonlinear incremental systems: Application to {PI}
  stabilization of nonlinear {RLC} circuits,'' \emph{{S}ystems \& {C}ontrol
  {L}etters}, vol.~56, no.~9, pp. 618--622, 2007.

\bibitem{maschke2000energy}
B.~Maschke, R.~Ortega, and A.~J. van~der Schaft, ``Energy-based {L}yapunov
  functions for forced {H}amiltonian systems with dissipation,'' \emph{IEEE
  Transactions on Automatic Control}, vol.~45, no.~8, pp. 1498--1502, 2000.

\bibitem{alonso2001stabilization}
A.~A. Alonso and B.~E. Ydstie, ``Stabilization of distributed systems using
  irreversible thermodynamics,'' \emph{Automatica}, vol.~37, no.~11, pp.
  1739--1755, 2001.

\bibitem{keenan1951availability}
J.~H. Keenan, ``Availability and irreversibility in thermodynamics,''
  \emph{British Journal of Applied Physics}, vol.~2, no.~7, p. 183, 1951.

\bibitem{wen2003unifying}
J.~T. Wen and M.~Arcak, ``A unifying passivity framework for network flow
  control,'' in \emph{INFOCOM 2003. Twenty-Second Annual Joint Conference of
  the IEEE Computer and Communications. IEEE Societies}, vol.~2.\hskip 1em plus
  0.5em minus 0.4em\relax IEEE, 2003, pp. 1156--1166.

\bibitem{de2017bregman}
C.~D. Persis and N.~Monshizadeh, ``Bregman storage functions for microgrid
  control,'' \emph{IEEE Transactions on Automatic Control}, vol.~PP, no.~99,
  pp. 1--1, 2017.

\bibitem{trip2016internal}
S.~Trip, M.~B{\"u}rger, and C.~De~Persis, ``An internal model approach to
  (optimal) frequency regulation in power grids with time-varying voltages,''
  \emph{Automatica}, vol.~64, pp. 240--253, 2016.

\bibitem{monshizadeh2017agreeing}
N.~Monshizadeh and C.~De~Persis, ``Agreeing in networks: Unmatched
  disturbances, algebraic constraints and optimality,'' \emph{Automatica},
  vol.~75, pp. 63--74, 2017.

\bibitem{JWAV:13}
J.~Wei and A.~van~der Schaft, ``Load balancing of dynamical distribution
  networks with flow constraints and unknown in/outflows,'' \emph{Systems \&
  Control Letters}, vol.~62, no.~11, pp. 1001--1008, 2013.

\bibitem{ferguson2017new}
J.~Ferguson, A.~Donaire, R.~Ortega, and R.~H. Middleton, ``New results on
  disturbance rejection for energy-shaping controlled port-{H}amiltonian
  systems,'' \emph{arXiv preprint arXiv:1710.06070}, 2017.

\bibitem{Hines2011}
G.~H. Hines, M.~Arcak, and A.~K. Packard, ``Equilibrium-independent passivity:
  {A} new definition and numerical certification,'' \emph{Automatica}, vol.~47,
  no.~9, pp. 1949 -- 1956, 2011.

\bibitem{Burger-duality2014}
M.~B{\"u}rger, D.~Zelazo, and F.~Allg{\"o}wer, ``Duality and network theory in
  passivity-based cooperative control,'' \emph{Automatica}, vol.~50, no.~8, pp.
  2051--2061, 2014.

\bibitem{simpson2017equilibrium}
J.~W. Simpson-Porco, ``Equilibrium-independent dissipativity with quadratic
  supply rates,'' \emph{arXiv preprint arXiv:1709.06986}, 2017.

\bibitem{van1995hamiltonian}
A.~van~der Schaft and B.~Maschke, ``The {H}amiltonian formulation of energy
  conserving physical systems with external ports,'' \emph{AEU. Archiv f{\"u}r
  Elektronik und {\"U}bertragungstechnik}, vol.~49, no. 5-6, pp. 362--371,
  1995.

\bibitem{Arjan2014PH}
A.~van~der Schaft and D.~Jeltsema, \emph{Port-{H}amiltonian Systems Theory: An
  Introductory Overview}.\hskip 1em plus 0.5em minus 0.4em\relax Now Publishers
  Incorporated, 2014.

\bibitem{schaft2013}
A.~van~der Schaft and B.~Maschke, ``Port-{H}amiltonian systems on graphs,''
  \emph{SIAM Journal on Control and Optimization}, vol.~51, no.~2, pp.
  906--937, 2013.

\bibitem{ferguson2015disturbance}
J.~Ferguson, R.~H. Middleton, and A.~Donaire, ``Disturbance rejection via
  control by interconnection of port-{H}amiltonian systems,'' in \emph{Decision
  and Control (CDC), 2015 IEEE 54th Annual Conference on}.\hskip 1em plus 0.5em
  minus 0.4em\relax IEEE, 2015, pp. 507--512.

\bibitem{bregman1967relaxation}
L.~M. Bregman, ``The relaxation method of finding the common point of convex
  sets and its application to the solution of problems in convex programming,''
  \emph{USSR Computational Mathematics and Mathematical Physics}, vol.~7,
  no.~3, pp. 200--217, 1967.

\bibitem{RO-motor-17}
R.~Ortega, N.~Monshizadeh, P.~Monshizadeh, D.~Bazylev, and A.~Pyrkin,
  ``Permanent magnet synchronous motors are globally asymptotically
  stabilizable with {PI} current control,'' 2017, submitted.

\bibitem{desoer2009feedback}
C.~A. Desoer and M.~Vidyasagar, \emph{Feedback Systems: {I}nput-Output
  Properties}.\hskip 1em plus 0.5em minus 0.4em\relax SIAM, 2009.

\bibitem{Arnold2013}
V.~I. Arnol'd, \emph{Mathematical Methods of Classical Mechanics}.\hskip 1em
  plus 0.5em minus 0.4em\relax Springer Science \& Business Media, 2013,
  vol.~60.

\bibitem{Nielsen2010}
F.~Nielsen, \emph{Legendre Transformation and Information Geometry}.\hskip 1em
  plus 0.5em minus 0.4em\relax CIG-MEMO2, 2010.

\bibitem{ryu2016primer}
E.~K. Ryu and S.~Boyd, ``Primer on monotone operator methods,'' \emph{Applied
  and Computational Mathematics}, vol.~15, no.~1, pp. 3--43, 2016.

\bibitem{nesterov2013introductory}
Y.~Nesterov, \emph{Introductory Lectures on Convex Optimization: A Basic
  Course}.\hskip 1em plus 0.5em minus 0.4em\relax Springer Science \& Business
  Media, 2013, vol.~87.

\bibitem{Shaik2013}
S.~Fiaz, D.~Zonetti, R.~Ortega, J.~Scherpen, and A.~van~der Schaft, ``A
  port-{H}amiltonian approach to power network modeling and analysis,''
  \emph{European Journal of Control}, vol.~19, no.~6, pp. 477--485, 2013.

\bibitem{caliskan2014compositional}
S.~Y. Caliskan and P.~Tabuada, ``Compositional transient stability analysis of
  multimachine power networks,'' \emph{IEEE Transactions on Control of Network
  Systems}, vol.~1, no.~1, pp. 4--14, 2014.

\bibitem{Arjan2016}
A.~van~der Schaft and T.~Stegink, ``Perspectives in modeling for control of
  power networks,'' \emph{Annual Reviews in Control}, vol.~41, pp. 119--132,
  2016.

\bibitem{forni2014differential}
F.~Forni and R.~Sepulchre, ``A differential {L}yapunov framework for
  contraction analysis,'' \emph{IEEE Transactions on Automatic Control},
  vol.~59, no.~3, pp. 614--628, 2014.

\bibitem{forni2013differential}
F.~Forni, R.~Sepulchre, and A.~J. van~der Schaft, ``On differential passivity
  of physical systems,'' in \emph{52nd IEEE Conference on Decision and
  Control}, Dec 2013, pp. 6580--6585.

\end{thebibliography}

\end{document}